# Davis-Greenstein alignment of non-spherical grains

A. Lazarian

Astronomy Department, University of Texas, Austin, TX 78712-1083

## ABSTRACT

Paramagnetic alignment of non-spherical dust grains rotating at thermal velocities is studied. The analytical solution is found for the alignment measure of oblate grains. Perturbative approach is used for solving the problem. It is shown that even the first approximation of the suggested iteration procedure provides the accuracy well within one percent of the expected measure of alignment. The results obtained are applicable both to paramagnetic and to superferromagnetic grains.

*Subject headings:* dust, extinction — ISM, clouds — ISM, polarization

## 1. Introduction

Discovered nearly half a century ago (Hiltner 1949, Hall 1949) the alignment of the ISM grains remains a long standing unsolved problem (see Goodman et al 1995). To solve this problem is not only challenging, but alluring, as this should enable one to use the wealth of polarimetric data for quantitative studies of the ISM.

Paramagnetic alignment of thermally rotating grains is known as the Davis-Greenstein process. Introduced as far back as 1951, this process was later criticized for not being strong enough (Spitzer 1978, Whittet 1992). To improve the mechanism its superparamagnetic and ferromagnetic modifications were introduced (Jones & Spitzer 1967, Duley 1978, Mathis 1986). At present, however, observational evidence for paramagnetic alignment is inconclusive (Hildebrand 1988) and we believe that one of the reasons for this unsatisfactory state of affairs has been the absence of the analytical theory of alignment for non-spherical grains. Indeed, alignment of non-spherical grains was studied through Monte-Carlo simulations (see Purcell 1969, Purcell & Spitzer 1971), through numerical solving of Langevin equation in Roberge, DeGraff & Flaherty (1993), but analytical results were obtained only for spherical grains (Jones & Spitzer 1967).

In present paper we address this problem and propose an analytical approach to studying alignment of non-spherical grains. First (section 2), we formulate the Fokker-Planck equation for non-spherical grains and find a small parameter inherent to the problem. Then we use a perturbative approach to solve the problem analytically. The recursive formula that we obtain enables one to calculate the measure of alignment with any requested degree of accuracy. However, we show, that even the first approximation is accurate within 1%, this value that should suffice present-day requirements. Further on (section 3) we show, that the assumption of superparamagnetism requires fine tuning of sizes of superparamagnetic particles, and therefore is improbable, but the assumption of superferromagnetic inclusions is more difficult to dismiss. The ferromagnetic modification of the theory corresponds to the two orders of magnitude increase in values of imaginary part of paramagnetic susceptibility and, as a result, superferromagnetic grains should be aligned in accordance with the analytical theory introduced in this paper.

If grain axis makes angle $\beta$ with magnetic field $\mathbf{H}$, for an ensemble of spheroidal grains the measure of alignment can be described by the Rayleigh reduction factor (Greenberg, 1968):

$$\sigma = \frac{3}{2}\langle\cos^2\beta - \frac{1}{3}\rangle, \tag{1}$$

where here and further on angular brackets $\langle\rangle$ denote an ensemble averaging. In the model of Roberge et al (1993) that we also adopt for the rest of the paper, it is assumed, that due to the Barnett relaxation, vector **J** is directed along the major axis of inertia[1], which for an oblate spheroid coincides with its short axis. Therefore within the adopted model the measure of **J** alignment coincides for oblate grains with the Rayleigh reduction factor.

Davis-Greenstein mechanism aligns grains rotating at thermal velocities. Such a rotation is expected for grains in molecular clouds, where the concentration of atomic hydrogen is low. Indeed, according to Purcell (1979), recoils from nascent $H_2$ molecules are likely to be the major driving force for suprathermal rotation. It is claimed in Lazarian (1995a) that aromatic carbonaceous & graphite grains do not attain suprathermal angular velocities due to $H_2$ formation. Therefore Davis-Greenstein process may also be applicable to some fraction of grains within the diffuse ISM.

## 2. Davis-Greenstein process

### 2.1. Diffusion coefficients for a model grain

For our present purposes, we will consider only oblate spheroidal grains as there are indications that aligned grains are oblate rather than prolate (Aitken *et al.* 1985, Lee & Draine 1985, Hildebrand 1988). The spheroid semiaxes parallel and perpendicular to the grain symmetry axis are denoted $a_m$ and $b_m$ $(b_m > a_m)$, respectively. The core-mantle interface is assumed to be a spheroid confocal with the mantle surface. The core semiaxes are denoted by $a_c$ and $b_c$. The eccentricity of the core $(i = c)$ as well as that of the mantle $(i = m)$ is

$$e_i = \sqrt{1 - \frac{a_i^2}{b_i^2}} \tag{2}$$

and may be different for different components of the grain.

It was shown in Jones & Spitzer (1967) that paramagnetic alignment can be described through the Fokker-Planck equation (see Reichl 1980)

$$\frac{\partial f}{\partial t} + \frac{\partial}{\partial J_i}(\langle \triangle J_i \rangle f) = \frac{1}{2}\frac{\partial^2}{\partial J_i \partial J_j}(\langle \triangle J_i \triangle J_j \rangle f), \tag{3}$$

---

[1] A quantitative discussion of Barnett relaxation in Lazarian(1994) showed that this approximation is true with high accuracy only when the grain rotational temperature substantially exceeds that corresponding to grain material.

where $f$ is a distribution function of the angular momentum $J$, while $\langle \triangle J_i \rangle$ and $\langle \triangle J_i \triangle J_j \rangle$ are coefficients. An important job of calculating the above coefficients was done in Roberge et al. (1993). Wherever possible we preserve the notations adopted in the said study to facilitate a comparison of our predictions with the numerical results obtained in Roberge et al. (1993). In fact, such a comparison was done in DeGraff, Roberge & Flaherty (1995) and proved an exellent agreement of the results obtained through these two approaches.

It is shown in Roberge et al. (1993) that only the tensor components

$$\langle (\triangle J_x)^2 \rangle = \frac{\sqrt{\pi}}{3} n m b_m^4 v_{th}^3 \left( 1 + \frac{T_d}{T_g} \right) [\Gamma_\perp (1 + \cos^2 \beta) + \Gamma_\parallel \sin^2 \beta], \tag{4}$$

$$\langle (\triangle J_y)^2 \rangle = \langle (\triangle J_x)^2 \rangle \tag{5}$$

and

$$\langle (\triangle J_z)^2 \rangle = \frac{2\sqrt{\pi}}{3} n m^2 b_m^4 v_{th}^3 \left( 1 + \frac{T_d}{T_g} \right) (\Gamma_\perp \sin^2 \beta + \Gamma_\parallel \cos^2 \beta) \tag{6}$$

are not equal to zero after accounting for the Larmour precession.[2] Above $T_d$ and $T_g$ are, respectively, dust and gas temperatures, $v_{th} = \sqrt{2kT_g/m}$ is the thermal velocity of gaseous atoms of concentration $n$ and mass $m$; the coefficients $\Gamma_\perp(e_m)$ and $\Gamma_\parallel(e_m)$ are geometrical factors

$$\Gamma_\perp(e_m) = \frac{3}{32}\{7 - e_m^2 + (1 - e_m^2)g_m(e_m) + (1 - 2e_m^2)[1 + e_m^{-2}(1 - [1 - e_m]^2 g(e_m))]\}, \tag{7}$$

$$\Gamma_\parallel(e_m) = \frac{3}{16}\{3 + 4(1 - e_m^2)g_m(e_m) - e_m^2[1 - (1 - e_m^2)^2 g(e_m)]\}, \tag{8}$$

with

$$g(e_m) = \frac{1}{2e_m} \ln \left( \frac{1 + e_m}{1 - e_m} \right). \tag{9}$$

(see Fig 1).

Note, that $\Gamma_\perp(e_m)$ and $\Gamma_\parallel(e_m)$, coincide for both spherical grains ($e_m \rightarrow 0$) when they attain the value of 1, and flakes ($e_m \rightarrow 1$), when they attain the value of 3/8. The fact that $\cos \beta$ enters the expression for the diffusion coefficients in Eqs (4) and (6) reflects the fact that grain disorientation due to gaseous bombardment depends on grain geometry.

As for the mean angular momentum increments, it was obtained in Roberge *et al.* (1993)

$$\langle \triangle J_i \rangle = -\frac{J_i}{t_{gas}}, \tag{10}$$

---

[2]This precession takes place at a scale $\sim 10^5$s if the Barnett effect endows a rotating grain with a magnetic moment. This time is substantially smaller than that of grain alignment.

where the subscript "i" stands for $x, y$ and $z$ while

$$t_{gas} = \frac{3}{4\sqrt{\pi}} \frac{I_{zz}^b}{nm b_m^4 v_{th} \Gamma_{\parallel}(e_m)}. \tag{11}$$

The diffusion coefficients discussed above characterize gas-grain interactions. The effect of magnetic field on grains is imprinted through the coefficients

$$\langle \triangle J_j \rangle_{mag} = -\frac{J_j}{t_{mag}}, \quad j = x, y \tag{12}$$

$$\langle \triangle J_z \rangle_{mag} = 0 \tag{13}$$

(see Jones & Spitzer 1967) and

$$t_{mag} = \frac{I_{zz}^b}{\kappa V B^2}, \tag{14}$$

where for slow rotation, $\kappa \approx 2.5 \cdot 10^{-12} T_d^{-1}$ s (Spitzer 1978). Similarly

$$\langle (\triangle J_j)^2 \rangle_{mag} = 2kT_d V B^2 \kappa, \quad j = x, y \tag{15}$$

and

$$\langle (\triangle J_z)^2 \rangle_{mag} = 0, \tag{16}$$

where Eqs (13) and (16) reflect the fact that magnetic field does not influence grains rotating about its direction.

## 2.2. Small parameter

It is a common knowledge that asymptotic analytical results are usually attainable if a small parameter is present in a model. The difficulty in direct solving the Fokker-Planck equation stems from the fact that $\langle (\triangle J_i)^2 \rangle$ depends on the direction of **J** in respect to the magnetic field. Jones & Spitzer (1967) did not face this problem as this dependence disappeared for spherical grains for which $\Gamma_{\parallel} = \Gamma_{\perp}$.

In general, $\Gamma_{\parallel}$ differs from $\Gamma_{\perp}$, but according to Fig. 2 their ratio is close to unity and therefore the parameter $\gamma$

$$\gamma = 1 - \frac{\Gamma_{\perp}}{\Gamma_{\parallel}} \tag{17}$$

does not exceed 0.2. We choose it to be the small parameter of our model. Thus Eqs. (4)-(6) may be rewritten as

$$\langle (\triangle J_x)^2 \rangle = A(1 - 0.5\gamma[1 + \cos^2 \beta]), \tag{18}$$

$$\langle (\triangle J_y)^2 \rangle = \langle (\triangle J_x)^2 \rangle, \tag{19}$$

$$\langle (\triangle J_z)^2 \rangle = A(1 - \gamma \sin^2 \beta), \tag{20}$$

where

$$A = \frac{\sqrt{\pi}}{3} n m^2 b_m^4 v_{th} \left(1 + \frac{T_d}{T_g}\right) \Gamma_\parallel. \tag{21}$$

The values of $\cos^2 \beta$ and $\sin^2 \beta$ depend on the attained degree of alignment, and therefore to find them, one needs to know the distribution function for grains. In our approach, we use the distribution function corresponding to $\gamma = 0$ to obtain the zero approximation $\cos^2 \beta_0$ and $\sin^2 \beta_0$. On the next step, the zero approximation is used to obtain the first approximation for the distribution function. This process may be continued until the required accuracy is attained.

## 2.3. Iteration procedure

We start with assuming $\gamma = 0$. In this case, the separation of variables is applicable to Eq. (3). The stationary equation for the $z$ component is as follows

$$\frac{1}{2} \frac{\partial^2}{\partial J_z^2} (\langle (\triangle J_z)^2 \rangle f_z) - \frac{\partial}{\partial J_z} (\langle \triangle J_z \rangle f_z) = 0, \tag{22}$$

which gives a formal solution

$$\ln f_z = -\frac{J_z^2}{t_{gas} \langle (\triangle J_z)^2 \rangle} + \mathrm{const}_1. \tag{23}$$

In the case of $x$ and $y$ components, one needs to account for paramagnetic relaxation and therefore

$$\ln f_j = -\frac{J_j^2}{\langle (\triangle J_j)^2 \rangle + \langle (\triangle J_j)^2 \rangle_{mag}} \left(\frac{1}{t_{gas}} + \frac{1}{t_{mag}}\right) + \mathrm{const}_2, \quad j = x, y \tag{24}$$

To characterize the relative importance of magnetic torque, it is advantageous to define

$$\delta_1 = \frac{\langle \triangle J_j \rangle_{mag}}{\langle \triangle J_j \rangle} = \frac{t_{gas}}{t_{mag}} = \frac{3}{4\sqrt{\pi}} \frac{\kappa V B^2}{n m v_{th} b_m^4 \Gamma_\parallel}. \tag{25}$$

With thus defined $\delta_1$ the problem is similar to that discussed in Jones & Spitzer (1967). We refer our reader to this work and write here the answer

$$\cos^2 \beta_0 = \frac{1}{1 - \mathrm{æ}_0^2} \left[1 - \frac{\mathrm{æ}_0}{\sqrt{1 - \mathrm{æ}_0^2}} \arcsin \sqrt{1 - \mathrm{æ}_0^2}\right], \tag{26}$$

where

$$\text{æ}_0^2 = \frac{1 + \delta_1 T_d / T_m}{1 + \delta_1} \tag{27}$$

and

$$T_m = \frac{1}{2}(T_d + T_g). \tag{28}$$

The difference of our result as compared to one in Jones & Spitzer (1967) is that we discuss alignment of non-spherical grains and therefore our $\delta_1$ is different from the corresponding parameter in the latter study; the temperature of grain rotation corresponds to a later modification of theory suggested in Purcell & Spitzer (1971) to account for thermolization of atoms over the grain surface. In fact, in the same paper we may find interesting qualitative considerations relevant to modification of $\delta_1$ for non-spherical grains. Surely, Purcell & Spitzer could not get the exact result, as at that time neither alignment of **J** due to the Barnett relaxation, not rapid Larmour precession due to the Barnett induced magnetic moment were apprehended, but we can only admire the scientific intuition of the two pioneers of grain alignment. The first approximation that we are going to discuss now accounts for the non-linearity of the Fokker-Planck equation.

We will use $\cos^2 \beta_0$ given by Eq. (26) instead of $\cos^2 \beta$ in Eqs. (18)-(20). The accuracy of the suggested approximation was confirmed by independent calculations in DeGraff et. al. (1995). The modified diffusion coefficients can be easily written down:

$$\left(\frac{1}{t_{gas}} + \frac{1}{t_{mag}}\right) = \frac{1}{t_{gas}}(1 + \delta_1) \tag{29}$$

and

$$t_{gas}\langle(\triangle J_j)^2\rangle = kT_g I_{zz}^b \left(1 + \frac{T_d}{T_g}\right)\left(1 - \frac{\gamma}{2}(1 + \cos^2 \beta_0)\right), \tag{30}$$

$$t_{gas}\langle(\triangle J_j)^2\rangle_{mag} = 2kT_d \delta_1 I_{zz}^b, \quad j = x, y, \tag{31}$$

while

$$t_{gas}\langle(\triangle J_z)^2\rangle = kT_g I_{zz}^b \left(1 + \frac{T_d}{T_g}\right)(1 - \gamma \sin^2 \beta_0). \tag{32}$$

Using Eq. (25), it is possible to write

$$t_{gas}(\langle(\triangle J_j)^2\rangle + \langle(\triangle J_j^2)\rangle_{mag}) = 2kI_{zz}^b\{T_m + T_d\delta_1\}\left[1 - \frac{\gamma(1 + \cos^2 \beta_0)T_m}{2(T_m + T_d\delta_1)}\right]. \tag{33}$$

Similarly

$$t_{gas}\langle(\triangle J_z)^2\rangle = kI_{zz}^b T_m(1 - \gamma \sin^2 \beta_0). \tag{34}$$

Substituting these coefficients into the Fokker-Planck equation (see Eq. 22) it is easy to obtain

$$\ln f_z = -\frac{J_z^2}{2kI_{zz}^b T_m(1 - \gamma \sin^2 \beta_0)} \tag{35}$$

and

$$\ln f_j = -\frac{J_j^2}{2kI_{zz}^b T_{av}[1 - 0.5\gamma(1 + \cos^2 \beta_0)]}, \quad j = x, y \tag{36}$$

where, following Jones & Spitzer (1967), we define

$$T_{av} = \frac{T_m + T_d \delta_1}{1 + \delta_1} \tag{37}$$

The distribution function

$$f = f_x f_y f_z = \text{const} \cdot \exp\left\{-\frac{J^2[1 - (1 - æ_1^2)\cos^2 \beta]}{2kI_{zz}^b T_{av}[1 - 0.5\gamma(1 + \cos^2 \beta_0)]}\right\}, \tag{38}$$

where

$$æ_1^2 = \frac{T_{av}}{T_m}\left\{1 - \gamma \frac{T_m(1 + \cos^2 \beta_0)}{2T_{av}(\delta_1 + 1)}\right\}\frac{1}{1 - \gamma \sin^2 \beta_0}. \tag{39}$$

With the adopted accuracy

$$æ_1^2 = æ_0^2(1 - \gamma q_0), \tag{40}$$

where

$$q_0 = \frac{1}{æ_0^2[1 + \delta_1]}[\cos^2 \beta_0 - 0.5\sin^2 \beta_0 - \{æ_0^2(1 + \delta_1) - 1\}\sin^2 \beta_0]. \tag{41}$$

The alignment is independent of $J$ amplitude. Therefore it is possible to perform the integration over all possible values of $J$ to obtain

$$W_1 = C_1(1 - [1 - æ_1^2]\cos^2 \beta)^{-3/2}, \tag{42}$$

where the constant $C_1 = 0.5æ_1$ is defined through normalization

$$C_1 \int_0^\pi \frac{\sin \beta \, d\beta}{(1 - [1 - æ_1^2]\cos^2 \beta)^{3/2}} = 1. \tag{43}$$

Using this angular distribution function it is possible to obtain the first approximation to $\cos^2 \beta$

$$\cos^2 \beta_1 = æ_1 \int_0^{\pi/2} \frac{\cos^2 \beta \sin \beta \, d\beta}{(1 - [1 - æ_1^2]\cos^2 \beta)^{3/2}}. \tag{44}$$

which provides

$$\cos^2 \beta_1 = \frac{1}{1 - æ_0^2(1 - \gamma q_0)}\left\{1 - \frac{æ_0\sqrt{1 - \gamma q_0}}{\sqrt{1 - æ_0^2(1 - \gamma q_0)}}\arcsin\sqrt{1 - æ_0^2(1 - \gamma q_0)}\right\}. \tag{45}$$

It is easy to see that similar equations relate "i" and "i+1" approximations. Indeed, for a given $\cos^2 \beta_i$, it is possible to show that

$$\cos^2 \beta_{i+1} = \frac{1}{1 - \mathscr{x}_0^2(1 - \gamma q_i)} \left\{ 1 - \frac{\mathscr{x}_0 \sqrt{1 - \gamma q_i}}{\sqrt{1 - \mathscr{x}_0^2(1 - \gamma q_i)}} \arcsin \sqrt{1 - \mathscr{x}_0^2(1 - \gamma q_i)} \right\}, \quad (46)$$

where

$$q_i = \frac{1}{\mathscr{x}_0^2(\delta_1 + 1)}[\cos^2 \beta_i - 0.5 \sin^2 \beta_i - \{\mathscr{x}_0^2(\delta_1 + 1) - 1\} \sin^2 \beta_i]. \quad (47)$$

Therefore any requested degree of accuracy can be ensured by a sufficient number of iterations in accordance with Eqs. (46)-(47). Further we will discuss the limiting cases of weak and strong alignment, where some simplifications are possible for the expressions obtained.

## 2.4. Limiting cases

The combination $1 - \mathscr{x}_0^2$ that enters Eqs. (45) and (46) can be rewritten as

$$1 - \mathscr{x}_0^2 = \delta_1 \frac{1 - T_d/T_m}{1 + \delta_1}. \quad (48)$$

Therefore for $T_d \ll T_m$ and $\delta_1 \gg 1$, the above expression tends to unity, while for either $(T_m - T_d)/T_m \ll 1$ or/and $\delta_1 \ll 1$, this expression tends to zero. The first case corresponds to strong alignment, the second to weak. Note, that for $(T_m - T_d)/T_m \ll 1$ our expressions not directly applicable as the Barnett relaxation is not efficient in such conditions (Lazarian 1994) and therefore the angular momentum deviates from the direction of the major axis of inertia.

*Strong alignment*

If $\mathscr{x}_0^2 \ll 1$, we may expand with the accuracy $O(\gamma^2)$ and $O(\gamma \mathscr{x}_0^3)$

$$\frac{1}{1 - \mathscr{x}_0^2(1 - \gamma q_0)} \approx \frac{1}{1 - \mathscr{x}_0} - \gamma \frac{\mathscr{x}_0^2 q_0}{(1 - \mathscr{x}_0^2)^2}, \quad (49)$$

$$\frac{1}{\sqrt{1 - \mathscr{x}_0^2(1 - \gamma q_0)}} \approx \frac{1}{\sqrt{1 - \mathscr{x}_0^2}} - \gamma \frac{\mathscr{x}_0^2 q_0}{2(1 - \mathscr{x}_0^2)^{3/2}}, \quad (50)$$

$$\arcsin \sqrt{1 - \mathscr{x}_0^2(1 - \gamma q_0)} \approx \arcsin \sqrt{1 - \mathscr{x}_0^2} - \gamma \frac{\mathscr{x}_0 q_0}{2(1 - \mathscr{x}_0^2)^{1/2}}. \quad (51)$$

Thus

$$\cos^2 \beta_1 \approx \frac{1}{1 - \ae_0^2} \left( 1 - \frac{\ae_0}{\sqrt{1 - \ae_0^2}} \arcsin \sqrt{1 - \ae_0^2} \right) + \gamma q_0 G(\ae_0), \tag{52}$$

where

$$G(\ae_0) \approx \frac{\pi}{4} \ae_0 - \frac{1}{2} \ae_0^2. \tag{53}$$

For small $\ae_0$, $G(\ae_0)$ tends to $0.25\pi\ae_0$. To determine the behavior of $q_0$ at this limit, we observe that

$$\ae_0^2(\delta_1 + 1) = 1 + \frac{T_d}{T_m}\delta_1. \tag{54}$$

Therefore for $T_d/T_m\delta_1 \gg 1$, $q_0 \approx (T_d/T_m\delta_1)^{-1} \ll 1$ and the order of the $\gamma$-term is smaller than in the case $T_d/T_m\delta_1 \ll 1$, $\delta_1 \gg 1$, when $q_0$ tends to unity. In the latter case the order of the $\gamma$-term is $\gamma\ae_0 \ll 1$. For $\ae_0$ close to unity, the $\gamma$-term should be compared with the measure of alignment which is of the order of unity, and this ensures a low value of a relative error even in the zero approximation.

*Weak alignment*

For $\ae_0 \to 1$, one can use a substitution

$$x^2 = 1 - \ae_0^2 \tag{55}$$

and expand the functions at $x = 0$. It is easy to see that

$$\begin{aligned}
\cos^2 \beta_1 &= \frac{1}{x^2} \left( 1 - \frac{\sqrt{1 - x^2}}{x} \arcsin x \right) \\
&\approx \frac{1}{x^2} \left( 1 - \frac{1}{x} \left[ 1 - \frac{1}{2}x^2 - \frac{1}{8}x^4 \right] \left[ x + \frac{x^3}{6} + \frac{3x^5}{40} \right] \right) \\
&= \frac{1}{3} + \frac{1}{10}x^2 = \frac{1}{3} + 0.1(1 - \ae_0^2) + 0.1\gamma\ae_0^2 q_0,
\end{aligned} \tag{56}$$

where $q_0 \approx 0.15(1 - \ae_0^2) = 2\delta_1 T_d/3T_m$, which is $0.15(1 - \ae_0^2)$ for $T_d \ll T_m$. Therefore the ratio of the $\gamma$-term to the combination $(\cos^2 \beta_1 - 1/3)$ that enters the measure of alignment is of the order $0.15\gamma \ll 1$, and this does not exceed 3% even for a maximal value of $\gamma$.

Therefore in both limiting cases of strong and weak alignment the $\gamma$-terms are within a few percent of the zero approximation.

To characterize the difference between the zero and the first approximation for arbitrary alignment, it is possible to introduce a function

$$\triangle = \frac{\sigma^{(1)} - \sigma^{(0)}}{\sigma^{(0)}}, \tag{57}$$

where $\sigma^{(1)}$ and $\sigma^{(0)}$ are the measures of alignment for the first and zero approximations, respectively. This function is shown in Fig. 3 for $\delta_1$ in the range from 1 to 10, $\gamma = 0.2$ and $T_m/T_a = 0.1$.

## 3. Enhancement of paramagnetic relaxation

Superparamagnetic grains were introduced in Jones & Spitzer (1967) in an attempt to account for observed alignment through Davis-Greenstein process. Superparamagnetic properties of grains emerge, according to Jones & Spitzer (1967), if grains contain ferromagnetic cluster within non-magnetic matrix. The susceptibility per volume becomes

$$\chi_{sp} = n\left(\frac{\mu^2}{3kT}\right) = \Upsilon\left(\frac{N\mu_0^2}{3kT}\right) = \Upsilon\chi_0 \tag{58}$$

(Jones & Spitzer 1967), where $n$ is the average number of magnetic precipitates per unit volume, $\mu$ is the magnetic moment, $N$ is the average number of ion atoms per unit volume, $\Upsilon$ is the average number of iron atoms per a precipitate particle, $\mu_0$ is the "effective" magnetic moment of each magnetic ion, $\chi_0$ is the ordinary paramagnetic susceptibility for a magnetic ion concentration $N$.

As the combination $N\mu_0^2/3k$ is the Curie constant, the susceptibility is enhanced by the factor $\Upsilon$ which is of the order $10^3 - 10^5$. The imaginary part of magnetic susceptibility is related to the static value of $\chi_{sp}(0)$ in the following way

$$\chi_{sp}'' = \chi_{sp}(0)\frac{\omega\tau}{1 + (\omega\tau)^2}, \tag{59}$$

where $\omega$ is the frequency of harmonic oscillations of the magnetizing field, and $\tau$ is the relaxation rate given by

$$\frac{1}{\tau} = \nu_0 \exp\left(-\frac{V_p \triangle E}{kT}\right), \tag{60}$$

where the factor $\nu_0$ represents the disorienting effects of random thermal modulations which non-magnetic matrix produces on a precipitate particle ($\nu_0 \sim 5 \cdot 10^9$ s$^{-1}$ for metallic iron precipitates), $V_p$ is their volume, and $\triangle E$ is the energy barrier restraining the magnetic reversal of individual particles ($\triangle E/k \sim 6.2 \cdot 10^{21}$ K cm$^{-3}$ (Brown 1959)).

If $\omega\tau \ll 1$, $\chi_{sp}'' \approx \chi_{sp}(0)(\omega\tau)$, while for $\omega\tau \gg 1$, $\chi_{sp}'' \approx \chi_{sp}(0)/(\omega\tau)$. Thus, $\omega\tau$ determines the relaxation. For a given $\omega$, the exponential dependence of $\tau$ determines a rather narrow range of $\Upsilon$ and this is, to our mind, a serious limitation for the suggested way of improving the alignment. For instance, Jones & Spitzer (1967) assume that $\omega = 10^5$ s$^{-1}$ and find that $V_p/T$ should be $\sim 1.8 \cdot 10^{-21}$ cm$^3$ K$^{-1}$, which for $T$ equal 10 K gives $1.8 \cdot 10^{-20}$ cm$^3$ K$^{-1}$. Therefore assuming that a unit-sell size for iron is $23.5 \cdot 10^{-24}$ cm$^3$ and $58.9 \cdot 10^{-24}$ cm$^3$ for magnetite, they obtain the the optimal number of atoms per precipitate particle $\Upsilon$ which is $1.5 \cdot 10^3$ and $7.4 \cdot 10^2$ atoms, respectively. Note that according to Eq. (58) this is the factor over which the susceptibility $\chi_{sp}$ exceeds $\chi_0$. Therefore $\kappa$ factor that enters Eq. (14) increase to $\leq 10^{-7}$ s$^{-1}$ for iron and magnetite precipitates as compared with $10^{-13}$ s$^{-1}$ for ordinary paramagnetic grains; the additional increase in imaginary part of susceptibility is due to the fact that this value was chosen to be maximal for frequency $10^6$ s$^{-1}$ for ordinary materials. This will be true for a very narrow range of $\Upsilon$. Indeed, a combination

$$\omega\tau = \omega\nu_0^{-1}\exp\left(\frac{V_p \triangle E}{kT}\right) \approx 5 \cdot 10^{14}\exp\left(\frac{V_p \triangle E}{kT}\right) \tag{61}$$

which controls $\chi_{sp}''$ changes drastically with $V_p$ or equivalently $\Upsilon$. In fact, less than 40% change in $\Upsilon$ is enough to reduce $\kappa$ values to those provided by ordinary paramagnetism. Provided that the alignment must be rather efficient ($\lesssim 20\%$ corresponds to the data in Serkowski et al (1975)), it is unlikely that a substantial fraction of grains contains precipitates of exactly required size. Similar criticism of superparamagnetic remedy for the alignment problem can be found in Duley (1978). We go further and emphasize that for thermally rotating grains, their velocity of rotation scales as $l^{-5/2}\varrho^{-1/2}T^{-1/2}$. Therefore we do not believe that the Nature fine-tunes to provide grains with precipitates of just required size to enable a substantial enhancement of paramagnetic relaxation for a wide range of grain shapes, sizes and densities.

This is even less probable for suprathermally rotating grains. The angular velocity of these grains, apart from shape, fractal dimension, and size, depends on the number of sites of $H_2$ formation, and on kinetic energy of $H_2$ molecules formed over particular types of sites (Purcell 1979, Lazarian 1995b). Moreover, tuning is required to be even more fine as for rotational velocities close to $10^9$ s$^{-1}$ and the paramagnetic susceptibility of ordinary grains reaches its maximal value; therefore it is only $\Upsilon$ factor of enhancement of $\kappa$ that we are speaking about.

Nevertheless, some enhancement of effective paramagnetic susceptibility is attainable for superferromagnetic grains. To understand why this is so, we note, that Davis-Greenstein mechanism is based on spin-spin relaxation and ignores the spin-lattice processes (see Jones & Spitzer 1967). According to Casimir & Du Pré (1938), the spin-lattice contribution to the susceptibility is given by

$$\chi_{sl} = \chi_0(H)\left[\frac{C_M}{C_H} + \left(\frac{C_H - C_M}{H}\right)\left(\frac{1}{1 + i\omega\tau_{sl}}\right)\right], \tag{62}$$

where $\chi_0(H)$ is the isothermal susceptibility at a bias field $H$, $C_M$ and $C_H$ are the magnetic heat at constant magnetization and the specific heat at constant field, respectively. The difference

$$C_H - C_M = T \left( \frac{\partial M}{\partial T} \right)_H^2 \left( \frac{\partial M}{\partial H} \right)_H^{-1} \tag{63}$$

(Caspers 1964), where $M$ is the magnetization of the material, becomes zero for the bias field $H$ equal zero and makes $\chi_{sl}''$ to vanish. In presence of static magnetic field (Morrish 1965)

$$\chi_{sl}'' = \left[ \frac{\omega \tau_{sl}}{1 + (\omega \tau_{sl})^2} \right] \frac{N \mu^2}{3kT} \frac{H_s^2}{H_s^2 + 0.5 H_i^2}, \tag{64}$$

where $\tau_{sl}$ is the spin-lattice relaxation time, $\mu$ is the Bohr magneton, $H_s$ and $H_i$ are, respectively, intensities of the static magnetic field and internal field in the paramagnetic material; the latter field is $\sim N \mu$.

Calculations in Duley (1978) for core-mantle grains provide an enhancement of $\chi_{sl}''$ up to a factor of $10^2$. This is much less that superparamagnetic enhancement could provide, but adequate for the purposes of alignment. However, a number of questions stay unanswered. Duley studied MgOFeOSiO cores which correspond to the grain models in Duley & Millar (1978). Since FeO is antiferromagnetic at least large clusters should show no net magnetic moment. Therefore Duley appeals to small particles of size less than 100 Å, that may in accordance with experiments by Richardson & Milligan (1956), Schuele & Deetscreek (1962) and Woods & Fine (1969) demonstrate ferromagnetic ordering. Therefore more efficient alignment should be present for the smallest grains, in particular, spin-lattice relaxation should occur in bare MgOFeOSiO particles, "since paramagnetic defect centers (lattice vacancies, H atoms, other paramagnetic ions) will experience the static field from $Fe^{+2}$ clusters. The resulting enhancement in power dissipation would permit 100-200 Å MgOFeOSiO cores to be aligned in microgauss fields" (Duley 1978). This tendency contradicts to observations which show a better alignment of large grains (see Kim & Martin 1994) and therefore the evidence in favor of ferromagnetic relaxation is inconclusive.

It is also important that the ferromagnetic relaxation provides rather limited enhancement of the imaginary part of magnetic susceptibility. Therefore grains must stay sensitive to magnetic field strength, which makes it possible to use results obtained in section 2 to calculate the corresponding measure of grain alignment.

## 4. Discussion

The iteration procedure introduced above enables one to calculate $\langle \cos^2 \beta \rangle$ and thus the measure of alignment with any required accuracy. However, we believe that for majority of practically interesting situations no iterations of order higher than one is needed. This is not only because a rather high accuracy attainable in the first approximation, but also due to more fundamental causes. We adopted the model by Roberge et al. (1993) in which **J** is directed along the axis of major inertia, and this assumption limits our accuracy. In fact, it is shown in Lazarian (1994) that the distribution of directions of **J** in the grain coordinates should be described by a Fokker-Planck equation, which solution is sharply peaked only when the rotational temperature of a grain is considerably different from that of the grain material. Therefore our solution is exact only in the limit of $T_d/T_g \to 0$. In our next paper (Lazarian, in preparation) we will derive analytics applicable to finite $T_d/T_g$.

Our approach presents several advantages as compared to direct numerical simulations. For instance, the formulae obtained above provide a clear physical insight into the phenomenon of paramagnetic alignment of interstellar grains; it shows that shapes of grains do effect grain alignment, but their influence comes mostly through $\delta_1$ factor.

By its own, Davis-Greenstein mechanism is not very much interesting for the vast regions of the ISM. And it is not only the problem with the magnetic field intensity, but mostly with the peculiarities of the polarization curve, which indicate more efficient alignment of large grains, while the Davis-Greenstein mechanism tends to go the other way round. However, if grains are superferromagnetic, the probability of having a ferromagnetic impurity increases with the grain size and the mechanism can be reconciled with the polarization curve (see Mathis 1986). The analytical solutions that we obtained, are applicable for describing alignment of such superferromagnetic grains.

This work would not be possible without friendly encouragement by Ethan Vishniac. I would like to thank the referee Wayne Roberge for a number of useful suggestions and for providing the comparison of my prediction with his numerical results. I am grateful to Bruce Draine, Lyman Spitzer and David Williams for illuminating discussions; valuable comments by John Mathis and Paul Shapiro are acknowledged. The manuscript was improved after my visit to the Princeton University Observatory, and I would like to thank Jeremiah Ostriker for providing the financial backing for the visit. The research is supported by NASA grant NAG5 2773.

---

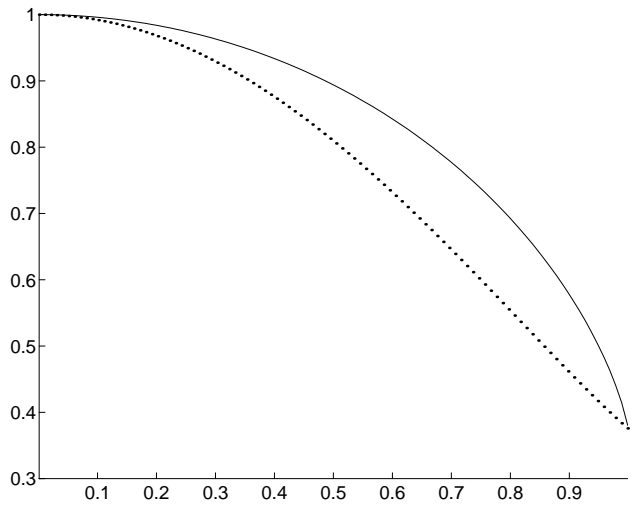

Fig. 1.— The values of $\Gamma_\perp$ (dotted line) and $\Gamma_\parallel$ (solid line) as a function of eccentricity $e_m$.

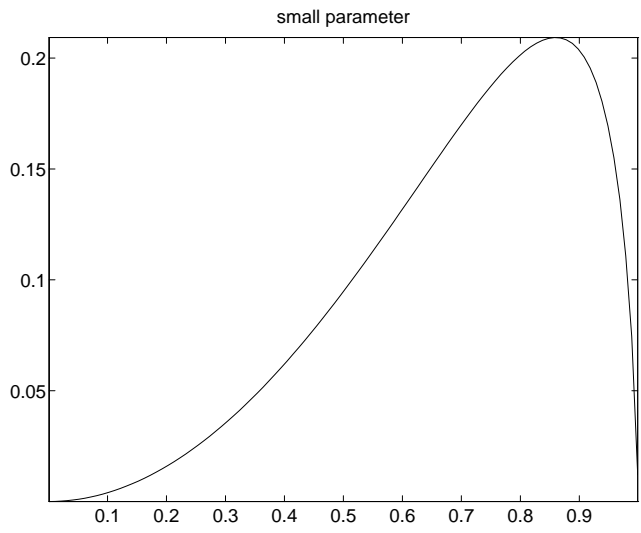

Fig. 2.— The small parameter of the problem as a function of grain eccentricity.

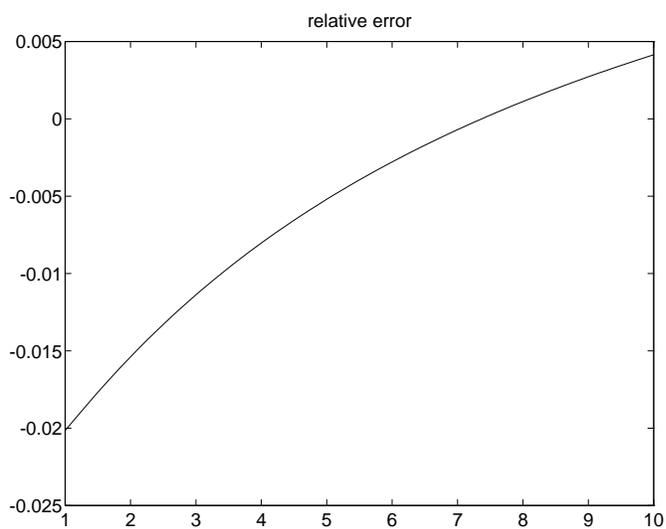

Fig. 3.— The relative difference in the alignment measures $\triangle$ for the zero and the first approximations as a function of $\delta_1$.